
\def\newline{\par}
\def\r#1{[#1]}
\null
\tolerance=10000
\baselineskip=24pt
{\it submitted to Physica Scripta 1994-11-07}\newline
{\it revised 1994-12-02}\hfill
\vskip 2cm
\centerline {\bf Measurement of isotope shift in Eu II}
\vskip 2cm
Lars Brostrom$^1$, Sven Mannervik$^1$, Peder Royen$^2$ and Anders
Wannstrom$^3$

{\it 1) Atomic Physics, Stockholm University, Frescativagen 24, S-104 05
Stockholm, Sweden\newline
2)Department of Physics, Stockholm University, Vanadisvagen 9, S-113 46
Stockholm, Sweden\newline
3)Department of Physics, Uppsala University, Box 530, S-751 21 Uppsala,
Sweden}
\vskip 2cm

{\bf Abstract}

The isotope shift between singly-charged ${\rm^{151}Eu}$ and ${\rm^{153}Eu}$
in the
$4f^7(^8S^o)6s~^9S_4^o~/--~4f^7(^8S^o)6p_{1\slash 2}~J=4$ transition at 4129 ]
has been measured using fast ion beam-laser technique. This Eu line has
attracted interest
in connection with efforts of obtaining a cosmochronometer based on
observed Th\slash Eu abundance ratios. Knowledge of the isotope shift is of
importance in order to check that contaminations from line blends do not
contribute to the line intensity of Eu II. The measured value of the isotope
shift -0.1527(2)${\rm cm^{-1}}$ is consistent with the old
spectroscopic value of Krebs and Winkler (-0.1503(25)${\rm cm^{-1}}$) using
a Fabry Perot interferometer,
while the accuracy is improved substantially.

\vfill
\eject

{\bf 1. Introduction}

Stellar spectra can be used for determination of elemental abundances.
Recently Butcher \r1 proposed that stellar spectra also could be used for
age determination of stars. The only long-lived isotope of thorium,
${\rm^{232}Th}$, has a half-life of 14 Gyr, which relates well to the time
scale
for stellar and galactic evolution. A useful cosmochronometer might be
obtained if the thorium abundance is compared to a stable element \r1.
Singly-charged thorium (Th II) has a useful spectral line at 4019.129 ].
Butcher \r1 chose a nearby transition in singly ionized neodymium (Nd II)
at 4018.823 ]
for comparison. He extracted abundance ratios of these spectral lines for
stars of very different ages and concluded that the galactic age is no more
than at maximum 9.6 Gyr, if a constant rate of nucleo-synthesis
throughout of the life of the Galaxy was assumed. For an exponentially decaying
production rate he obtained 11-12 Gyr.

The proposed method involves some problems regarding both Th and Nd. The Th
II line is blended with a Co I line \r2. The cobalt line is much more sensitive
to temperature which causes the contribution from cobalt to vary very much
from star to star. Lawler et al \r2 have analyzed the contamination from
cobalt in the Th II intensities of the same set of stars that was analyzed
by Butcher. From the revised analysis of the spectra and in comparison with
predictions based on models of chemical evolution,
they  suggested an age of the Galaxy of 15-20 Gyr. A more detailed
analysis, utilizing spectrum synthesis, by Morell et al \r3 revealed the
need to introduce further blends in the spectrum close to the position of
the Th II line in order to reproduce the spectrum. They conclude that due
to the scattering in the Th\slash Nd abundance ratio, it is not possible to
draw any firm conclusion about the age of the Galaxy. A further
complication regarding the Th II line was discussed recently by  Grevesse and
Noels \r4. They found that the solar Th abundance derived from the 4019 ] line
intensity was 40/
that this discrepancy could be explained and corrected for by a blend of
the V I intercombination line $a~^2D_{5\slash 2} - w~^4P_{5\slash 2}$.

There are also problems with the use of Nd as a stable
element reference. The thorium isotope is only created by the r-process
(rapid neutron capture), whereas neodymium  only partly is formed by this
process, the rest being created by the s-process (slow neutron capture).
Although Butcher \r1 claims that the partial contributions do not vary
substantially from star to star, an alternative stable element has been
searched for. As suggested by Pagel \r5, europium is a good candidate,
almost entirely formed by the
r-process. A Eu II line at 4129.70 ] seems to be a good trace of stellar
europium abundance. The line appears to be free from line blends (and its
wavelength is not too far from that of
the Th II line). Thus, it has been suggested that this line
should be compared with the Th II intensity.  Recently, Fran\c cois et al \r6
have
investigated the Th\slash Eu ratios in a set of metal poor stars.
They obtained a
complex variation in the actual ratio, from which no conclusions could be
drawn regarding the age of the Galaxy. It is not clear from this paper on
which atomic data the analysis was relying. We presume that the Eu II line
mentioned here (4129.7 ]) was used by Fran\c cois et al and that the line 4013
]
mentioned in their article was a misprint. It was not mentioned whether
hyperfine structure and isotope shift were considered in the analysis.

It is important that the 4129.7 ] Eu II line in the solar spectrum,
with known hyperfine structure and isotope shift for
the europium isotopes, can be reproduced otherwise blends must be searched for.
A weak blend in the solar spectrum
might be strong in other types of stars, especially if the blend comes from a
neutral element of the iron group. For this reason it is of great
importance to have accurate values for hyperfine structure and isotope
shift in order to be able to synthesize a very accurate line profile from
which it could be concluded whether or not a blend is present.

The hyperfine structure of the Eu II
$4f^7(^8S^o)6s~^9S_4^o~-~4f^7(^8S^o)6p_{1\slash 2}~J=4$ transition at
4129.70 ] has recently been accurately determined in the two isotopes
${\rm ^{151}Eu}$ and ${\rm ^{153}Eu}$. Becker et al \r7 have
measured the hyperfine splitting of the ground state
($4f^7(^8S^o)6s~^9S_4^o$) using laser rf double-resonance technique in a Paul
trap. By this method they have been able to determine the hyperfine
splittings, which are in the range 1-10 GHz, with an accuracy of the order
of 100 Hz. The hyperfine structure of the upper level has been determined
by Villemoes and Wang \r8 using the fast ion beam-laser method. Here the
splittings are of the order 0.1-1 GHz, and these have been determined with
an accuracy of about
1 MHz. Thus, there are reliable hyperfine data which can be used in the
analysis of stellar spectra.

For the isotope shift there is an old spectroscopic measurement by Krebs
and Winkler \r9 using a Fabry-Perot interferometer, which partially could
resolve the hyperfine components of the spectrum. Krebs and Winkler
obtained an isotope
shift of -0.1503${\rm \pm 0.0025~cm^{-1}}$ for this particular transition.
However,
due to the efforts of finding a cosmochronometer there has been a request
from astrophysicists \r{10} to remeasure this isotope shift with the accurate
beam-laser technique in order to check the old measurement and to firmly
establish the isotope shift for this transition. Thus we decided to perform
a measurement of this quantity.

{\bf 2. Experimental set-up}

At the CRYRING facility of the Manne Siegbahn Laboratory of Stockholm
University there is an ion source CRYSIS which has an ion injector (INIS),
which apart from its function as injector also can be used
separately for fast ion beam-laser spectroscopy. INIS is an isotope
separator usually operated at 25 kV. The experimental set-up has been
described in some detail previously \r{11}. The ions were produced in a
low-voltage
electron-impact ion source. Europium was introduced in metallic form and
the arc in the ion source was supported by letting a noble gas into the
discharge region. The two isotopes of europium (${\rm^{151}Eu}$ and
${\rm^{153}Eu}$) are easily separated by the mass analyzing magnet.

An ion current of a few hundred nA was collected for each isotope (the two
isotopes have almost the same abundance) in a Faraday cup, which can be
inserted in front of the interaction chamber. With a system of
electrostatic deflection plates and quadrupoles the beam could be
transported into the interaction region.

The laser light was introduced into
the vacuum system in antiparallel direction with respect to the ion beam
(Fig 1). To assure that the ion beam  and the laser beam travel along the
same path in the interaction chamber, 4 mm diameter apertures are mounted on
each side of the interaction chamber (at a distance of 1.4 m from each other).
Inside the chamber the ion beam velocity is locally changed by a Doppler Tuning
Device (DTD), consisting of three electrically isolated cylindrical tubes.
The center tube is connected to high voltage while the others are grounded.
The DTD was located in the focus of the optical fluorescence detecting
system. Thus, due to the local change of the Doppler shift, the laser
excitation
could be restricted to the region where fluorescence could be detected.
This is essential in order to avoid reduction of signal intensity by
optical pumping of the ion beam upstreams the detection region.

Through a slot in the center tube a lens system collected fluorescence light
and focused it on the entrance slit of a 0.35 m Heath EUE-700 monochromator
(Czerny-Turner mounting, f\slash 6.8). With slits set at 2 mm, the bandwidth
was 40 ]. Photons were counted by a EMI 9789 QA photomultiplier, which was
cooled by a Hamamatsu Peltier cooler. More details about the experimental
set-up and also the use of the DTD for lifetime measurements can be found
in our previous papers \r{11-13}.

A single-mode Coherent 699-29 Autoscan dye laser (with Stilbene 1) was
used to provide tunable laser light at 4129.7 ] (in the rest frame of the
ion) with a bandwidth of less than 1 MHz. The ring dye
laser was pumped by the UV lines of an Innova 400-20 argon ion laser.
About 100 mW tunable single frequency blue laser light
was obtained. This laser power, however, turned out to be too low to
support the Autoscan function of the ring dye laser. The laser was instead
operated in the 699-21 mode, i.e. utilizing the 30 GHz continuous scan
option. Since the laser power was too low (in combination with the high
laser frequency) to permit use of the internal wavemeter, an external
Burleigh  WA-4500 wavemeter was used to find the initial laser
frequency of the 30 GHz scan. The wavemeter has an accuracy of 0.02 ${\rm
cm^{-1}}$. During the laser scan the wavemeter was monitored in
order to observe whether mode hops occurred.
\vfill
\eject
{\bf 3. Experiment}

The goal of the present experiment was to measure the isotope shift in the
$4f^7(^8S^o)6s~^9S_4^o~/--~4f^7(^8S^o)6p_{1\slash 2}~J=4$ transition for
reasons mentioned above. This transition occurs at 4129.70 ], i.e. 24207.86
\r{14} ${\rm cm^{-1}}$. For excitation in
antiparallel geometry, the wavenumber of
the exciting laser light has to be reduced by about 15 ${\rm cm^{-1}}$ to
compensate for the Doppler shift. Fluorescence was observed by monitoring the
$4f^7(^8S^o)6s~^7S_3^o~-~4f^7(^8S^o)6p_{1\slash 2}~J=4$ transition at
4435.55 ] (22539 ${\rm cm^{-1}}$).

The basic idea of the experiment was to
switch isotopes during the laser scan in order to relate the hyperfine
components of the two isotopes to the same laser frequency scale, without
requirements of a very accurate knowledge of the laser frequency on an
absolute scale \r{15}. Consequently the
isotope shift will be exposed directly in the spectrum. In the analysis,
however, the different Doppler shifts of the two isotopes must be taken
into account. Since both isotopes were accelerated by the same high
voltage, the velocties will differ due to the isotopes different masses.
This Doppler `isotope shift' is of the same order of magnitude as the real
isotope shift.

In Fig 2a simulated spectra are shown, which are based on the
hyperfine splittings given in refs \r7 and \r8 (with a resolution similar to
that of the present experiment). The simulated spectrum of ${\rm^{151}Eu}$ is
displaced relative that of ${\rm^{153}Eu}$ due to the isotope
shift as well as due to the different Doppler shifts as indicated
schematically in the figure. Each spectrum
reflects the different splittings of the upper and the lower level. The
large splittings between the lower levels give rise to the large distances
between
six groups of levels according to Becker et al \r7. The distance between lines
inside such a spectral line group reflects the hyperfine splitting of the
upper level, as given by Villemoes and Wang \r8. In the
present experiment a few survey spectra were recorded to identify the
different components of the spectra. In Fig 2b such an example is given.

For the isotope shift measurement we decided to record one three-line group
of hyperfine components in ${\rm^{151}Eu~({11\over2}-{9\over2},~
{11\over2}-{11\over2},~{11\over2}-{13\over2})}$ (where $F-F^\prime$
denotes the hyperfine quantum number for the lower and the upper level,
respectively), then switch isotope to ${\rm^{153}Eu}$
and record the three-line group $({9\over2}-{7\over2},~{9\over2}-{9\over2},
{}~{9\over2}-{11\over2})$ and then again change back to isotope ${\rm^{151}Eu}$
and record the three-line group $({9\over2}-{7\over2},
{}~{9\over2}-{9\over2},~{9\over2}-{11\over2})$; all these nine lines recorded
in one single laser scan. The isotope separator was operated by a computer
based control
system. This system permitted programming of the magnet steering so that
isotopes could be shifted by a single strike of a key (the isotopes
shifted in about one second). Several spectra were recorded by this
method.

Since the
frequencies of the hyperfine splittings are well-known, the spectra could
be internally calibrated. The six ${\rm^{151}Eu}$ lines were used for this
purpose. The frequency scale was fitted to a third-degree polynomial. As
mentioned above the external wavemeter was only used to determine the frequency
of the starting point of the laser scan and to check that no mode hops
occurred during the scan. The fluorescence photon pulses were amplified,
discriminated and read into a CAMAC scaler. A PC was used to read the
scaler and store the value in a multi-scaler program as the laser frequency
was scanned. A typical spectrum
(Fig 3) was recorded in 500 s (1 second per point).

The spectral line positions were determined by a peak-fitting program. To
obtain as high intensity as possible in the spectrum, fluorescence was
observed close to the field gap of the DTD. This caused the lines to
be slightly asymmetric, since in the high frequency wing of the line,
fluorescence from excitation in the gap will contribute. In the present
experiment the voltage on the DTD was -401 V, yielding a higher beam
velocity inside the DTD than outside. In the fitting program the
lines were approximated by a Gaussian line profile with an exponential foot
on the high frequency wing (to take care of the asymmetry). The Gaussian
width reflects the velocity spread in the ion beam. In the present
experiment a typical Gaussian width of about 180 MHz (FWHM) was obtained.
To obtain the isotope shifts, the experimental hyperfine constants \r{7,8}
were used to determine the centre-of-gravity of the multiplets.
The weighted average of the `apparent'
isotope shift (i.e. the shift in the recorded spectrum without Doppler
correction) could be determined with an accuracy of 4 MHz, where
statistical errors both from the calibration procedure and the
determination of the peak positions are included.

After the line positions were determined, correction for the different
Doppler shifts of the two isotopes had to be done. The beam velocity can be
accurately determined if the ion beam is excited both in parallel and
antiparallel geometry \r{16} according to the formula

$$\beta ={v\over c} = {{\sigma_+-\sigma_-}\over {\sigma_++\sigma_-}}$$

,where $\sigma_+$ is the laser frequency for excitation in parallel
configuration and $\sigma_-$ for antiparallel. However, due to the
practical circumstances, it was more
reliable to apply this technique to a Xe II transition, for which
the laser dye Rhodamine 6G could be used. The laser system could thus be
run in a more stable mode utilizing the Autoscan option and the use of
the internal wave meter. The xenon transition is also stronger and the
single line appearing in ${\rm ^{132}Xe}$ is unambiguous. By this method the
high voltage could be accurately be determined, which could be used to
calculate the velocity of the europium isotopes, which were accelerated by
the same high voltage.

The high voltage calibration was performed by exciting the Xe II transition
$5d~^4D_{7\slash2}~-~6p~^4P^o_{5\slash2}$ at 6051.15] (16521.22 ${\rm
cm^{-1}}$) \r{17} (in the ion rest frame) and observing the laser-induced
fluorescence at 5292 ] ($6s~^4P_{5\slash2}~-~6p~^4P^o_{5\slash2}$),
as the laser frequency was tuned. Iodine lines \r{18} were used as reference
lines. Simultaneously, the Vernier etalon signal (of the internal wave
meter) as well as the transmission of an external high finesse Fabry-Perot
interferometer (2 GHz FSR) were recorded. Statistics of 16 scans for
$\sigma_-$ and 10 scans for $\sigma_+$ were used
in this calibration procedure and the high
voltage was determined to be 24454 ${\rm \pm}$ 10 V.
In the calibration
procedure the DTD was used in the same position and at the same voltage as
in the europium run. Thus it did not matter that the beam was not observed
in a completely field free region in the DTD (as found through the calibration
measurements). From the calibration of the high voltage the velocities for
the europium isotopes could be determined. For -401 V on the DTD
$\beta (^{151}Eu)=(5.9405\pm 0.0012)\cdot 10^{-4}$ and
$\beta (^{153}Eu)=(5.9015\pm 0.0012)\cdot 10^{-4}$ were found. These
$\beta$-values were used to correct for the Doppler shifts.
Since both isotopes were recorded in the same laser scan, it is only the
difference in velocity for the two isotopes which matters which means that
this determination is not so critical on an absolute scale. The velocity
difference is determined directly from the mass difference,
since both isotopes are accelerated by the same high voltage.
In the non-relatistic limit
$\beta (^{153}Eu)=\sqrt {m(^{151}Eu)\slash m(^{153}Eu)}\cdot\beta
(^{151}Eu)$, which implies that the uncertainty in the isotope shift
due to the uncertainty in the difference in velocity ($\delta(\Delta\beta)$)
is only
$\Delta\beta\lbrack 1-\sqrt{m(^{151}Eu)\slash m(^{153}Eu)} \rbrack\nu\approx0.5
{}~{\rm MHz}$ (where $\Delta\beta$ is the uncertainty in $\beta$ as given above
and $\nu$ is the actual laser frequency). All calculations were, however,
performed using the full relativistic expressions for
beam energies and Doppler shifts. Also for the internal frequency
calibration mentioned above, full relativistic Doppler correction was made.
Atomic masses ($m(^{151}Eu)$ and
$m(^{153}Eu)$) were obtained from the mass
tables of Wapstra and Bos \r{19}.

{\bf 4. Results and discussion}

{}From the evaluation of the measurements described above an isotope shift of
- 152.7 ${\rm\pm}$ 0.2 mK (i.e. -0.1527 ${\rm cm^{-1}}$ or 4578 MHz)
was obtained. The present result coincides within error bars with the
value of Krebs and Winkler \r9, who obtained an isotope shift of
-150.3${\rm\pm}$2.5 mK for this particular transition. It is impressive
that the old classical work could give such an accurate result though the
different hyperfine components were not very well resolved.

With the present result for the isotope shift and the recent accurate
values for the hyperfine structure \r{7,8}, there is now a solid ground of
atomic data which can be used in the analysis of europium in stellar spectra.

In the interpretation of stellar spectra more caution might be needed due
to the recent observations of isotopic abundance anomalies (see e.g.
\r{20}). The relative intensity between the ${\rm ^{151}Eu}$ and
the ${\rm ^{153}Eu}$
lines might not be a universal constant for different objects.

\vskip 2cm
{\bf Acknowledgment}

We would like to thank Dr N Grevesse for initiating this investigation
and for valuable discussions. Clarifying discussions with Dr O Morell is
also gratefully acknowledged.
We are grateful to the staff of the CRYRING
facility of the Manne Siegbahn Laboratory for providing accelerator time
and technical support. We would also like to thank Prof L Klynning, who put
the argon ion laser to our disposal.

This work was supported by grants from the Swedish Natural Science Research
Council (NFR).

\vskip 2cm
{\bf References}

1. H.R. Butcher, Nature {\bf 328}, 127 (1987). \newline
2. J.E. Lawler, W. Whaling and N. Grevesse, Nature {\bf 346}, 635
(1990).\newline
3. O. Morell, D. K{llander and H.R. Butcher, Astron. Astrophys. {\bf 259},
543  (1992).\newline
4. N. Grevesse and A. Noels, Physica Scripta {T47}, 133 (1993).\newline
5. B.E.J. Pagel, in `Evolutionary Phenomena in Galaxies' (1989), eds J.
Beckman, B.E.J. Pagel, Cambridge University Press, p 201.\newline
6. P. Fran\c cois, M. Spite and F. Spite, Astron. Astrophys. {\bf 274},
821 (1993). \newline
7. O. Becker, K. Enders, G. Werth and J. Dembczynski, Phys Rev {\bf A48},
3546 (1993).\newline
8. P. Villemoes and M. Wang, Z Phys D {\bf 30}, 19 (1994).\newline
9. K. Krebs and R. Winkler, Z Phys {\bf 160}, 320 (1960).\newline
10. N. Grevesse, private communication (1993).\newline
11. L. Brostrom, S. Mannervik, A. Passian and G. Sundstrom, Phys Rev {\bf
A49}, 3333 (1994).\newline
12. R.T. Short, S. Mannervik, M. Larsson, P. Sigray and D. Sonnek, J Phys
B{\bf 22}, L27 (1989).\newline
13. R.T. Short, S. Mannervik, M. Larsson, P. Sigray and D. Sonnek, Phys Rev
{\bf A39}, 3969 (1989).\newline
14. W.C. Martin, R. Zalubas and L. Hagan, `Atomic Energy Levels - The
Rare-Earth Elements', Nat. Bur. Stand. NSRDS-NBS 60 (1978) p 201.\newline
15. P. Villemoes, A. Arnesen, F. Heijkenskjold and A. W{nnstrom, J. Phys.
B{\bf 26}, 4289 (1993).\newline
16. O. Poulsen, Nucl Instr and Meth {\bf 202}, 503 (1982). \newline
17. J.E. Hansen and W. Persson, Physica Scripta {\bf 36}, 602 (1987).\newline
18. S. Gerstenkorn and P. Luc, `Atlas du spectre d'absorption de la
molecules d'iode 14800-20000 ${\rm cm^{-1}}$', ed du CNRS, Paris
(undated).\newline
19. A.H. Wapstra and K. Bos, Atomic Data and Nuclear Data Tables {\bf 19},
175 (1977). \newline
20. D.S. Leckrone, S. Johansson, G.M. Wahlgren and S.J. Adelman, Physica
Scripta {\bf T47} 149 (1993).\newline

\vfill
\eject
FIGURE CAPTIONS

Fig 1\newline
Schematic figure of the experimental set-up used for the isotope shift
measurements in europium.

Fig 2\newline
a. Simulated spectra of ${\rm ^{151}Eu~and~^{153}Eu}$ using the hyperfine
constants given by Becker et al \r7 and Villemoes and Wang \r8. The spectra
are shifted relative each other according to the different Doppler shifts
at the beam energy used in the present experiment. The figure illustrates
how the centres-of-gravity of the multiplets shift due to the isotope shift
and the difference in Doppler shift to give the observed shift (which is
denoted by the `apparent' isotope shift in the text).
\newline
b. An experimental survey spectrum covering 14 of the 16 components of the
$4f^7(^8S^o)6s~^9S_4^o~/--~4f^7(^8S^o)6p_{1\slash 2}~J=4$ transition in
${\rm ^{153}Eu}$.

Fig 3\newline
One of the experimental spectra used for extracting the isotope shift
between ${\rm ^{151}Eu~and~^{153}Eu}$ in the
$4f^7(^8S^o)6s~^9S_4^o~/--~4f^7(^8S^o)6p_{1\slash 2}~J=4$ transition in Eu II.
The spectrum shows detected laser-induced fluorescence from the ion beam as the
frequency of the ring dye laser was continuously scanned. During the laser
scan the isotope in the ion beam was changed. The spectrum starts with
fluorescene from ${\rm ^{151}Eu^+}$. When the three first components had
been recorded, the setting of the magnet of isotope separator was instantly
switched to give ${\rm ^{153}Eu^+}$. Three components were recorded in this
isotope. Then the beam was switched back to the ${\rm ^{151}Eu^+}$ isotope and
three new components were recorded in this isotope. This procedure permits
internal calibration of the frequency scale using the accurate hyperfine
splitting reported by Becker et al \r7.
\vfill
\eject
\end